\documentclass[fleqn]{SCGE2020}
\usepackage{booktabs}
\usepackage{hyperref}  
\usepackage{graphicx}
\usepackage{float} 

\newcommand{\beq}{\begin{equation}}
\newcommand{\eeq}{\end{equation}}
\newcommand\lsim{\mathrel{\rlap{\lower4pt\hbox{\hskip0pt$\sim$}}
        \raise1pt\hbox{$<$}}}
\newcommand\gsim{\mathrel{\rlap{\lower4pt\hbox{\hskip0pt$\sim$}}
        \raise1pt\hbox{$>$}}}

\begin{document}
 
\ensubject{subject}


\ArticleType{Article}
\SpecialTopic{SPECIAL TOPIC: }
\Year{2023}
\Month{January}
\Vol{66}
\No{1}
\DOI{??}
\ArtNo{000000}
\ReceiveDate{January 11, 2023}
\AcceptDate{April 6, 2023}

\title{Hierarchical Parameter Estimation for Intrinsic Parameters of Extreme-mass-ratio Inspirals Using Deep Learning}{Hierarchical Parameter Estimation for Intrinsic Parameters of Extreme-mass-ratio Inspirals Using Deep Learning}

\author[1,2]{Ping Shen}{}
\author[2,3,4,5]{Wen-Biao Han}{wbhan@shao.ac.cn}
\author[6]{Minghui Du}{}

\AuthorMark{P. Shen}%
\AuthorCitation{P. Shen, W.-B. Han, and M.-H. Du}


\address[1]{Faculty of Electronic Information Engineering, Huai'an University, Huaian {\rm  223003}, China}
\address[2]{Shanghai Astronomical Observatory, Chinese Academy of Sciences, Shanghai {\rm 200030}, China}
\address[3]{School of Fundamental Physics and Mathematical Sciences, Hangzhou Institute for Advanced Study, UCAS, Hangzhou {\rm 310024}, China} 
\address[4]{Key Laboratory of Radio Astronomy and Technology, Chinese Academy of Sciences, A20 Datun Road, Chaoyang District, Beijing {\rm 100101}, China} 
\address[5]{Shanghai Frontiers Science Center for  Gravitational Wave Detection, 800 Dongchuan Road, Shanghai {\rm 200240}, China} 
\address[6]{Center for Gravitational Wave Experiment, National Microgravity Laboratory, Institute of Mechanics, Chinese Academy of Sciences,
Beijing {\rm100190}, China} 

\abstract{

Extreme-mass-ratio inspirals (EMRIs) stand as prime targets for future space-borne gravitational wave (GW) detectors, yet their precise parameter estimation is severely hindered by waveform complexity and high-dimensional parameter spaces. In this work, we propose a deep learning framework capable of recovering the seven intrinsic EMRI parameters without relying on detailed priors for extrinsic parameters. Our approach first employs a U-Net to extract robust signal features from noisy time-frequency data. These features are then fed into a multi-channel convolutional neural network (CNN) equipped with a dual attention mechanism, which decouples the time and frequency dimensions to jointly model long-range temporal evolution and persistent spectral structures, thereby enhancing sensitivity to EMRI time–frequency ridges in noisy data.
Remarkably, this method reduces the search ranges for the parameters $\{\log M, a, \mu, e_0, p_0 \}$ and $\{t_p, {x_I}_0 \}$ by factors of $\sim 10^{2}$ and $\sim 10^{1}$, respectively, effectively shrinking the total volume of the seven-dimensional parameter space by $\sim 10^{8}$.  Such a dramatic reduction enables a rapid pre-screening of the intrinsic parameter space, substantially saving computational cost for subsequent full parameter estimation. Although our framework is developed with Taiji mission specifications in mind, the methodology is applicable to data from other space-borne GW detectors.

}


\maketitle


\begin{multicols}{2}

\section{Introduction}\label{intro}

Extreme-mass-ratio inspirals (EMRIs)~\cite{2007EMRI_IMRI,2018EMRI,vu2026selfforcecalculationsnumericalrelativity,xue2026recoilregulatedextrememassratioinspirals,black2026extrememassratioinspirals,zhang2026detectabilityextrememassratio,lui2026pitchingcosmiccurveballsenvironmental,zhou2026noteqpetimingfalse,jiangye} are among the primary targets for future space-borne gravitational-wave (GW) detectors, such as Laser Interferometer Space Antenna (LISA)~\cite{2022LISA,2017LISA,2019LISA}, Taiji~\cite{2020Taiji,2026TaijiChallenge_DMH,shen_2026_BHseed,zhongxingyu}, and TianQin~\cite{2016TianQin,2021TianQin,EMRI_TianQin2021_ZTG}. An EMRI consists of a stellar-mass compact object  inspiralling into a supermassive black hole (SMBH). Owing to the
\Authorfootnote 
\noindent 
 highly asymmetric masses of the two components, the mass ratio of these systems typically ranges from $10^{-4}$ to $10^{-7}$, 
which gives rise to the term ``extreme-mass-ratio inspiral''.

EMRIs carry rich scientific information with broad applications in astrophysics, cosmology, and fundamental physics. In astrophysics, their signals encode key source properties such as black hole mass, spin, orbital eccentricity, and inclination, enabling detailed studies of compact-object systems in galactic centers and their formation history~\cite{2026EMRI_e,2023LISA_astro}. In cosmology, if electromagnetic counterparts are identified, EMRIs can serve as “standard sirens” to measure luminosity distance and redshift, thereby constraining cosmological parameters such as the Hubble constant~\cite{2026EMRI_Hubble_constant,2023Cosmology_LISA,2025_LJD_EMRI_Hubble_Constant}. From the perspective of fundamental physics, their long-duration, multi-cycle evolution in the strong-field regime makes EMRIs highly sensitive to spacetime geometry, providing a powerful probe of general relativity (GR)~\cite{shen2023_massRatio,shen_2025_bumpy,2022_fundamental_LISA,2025_ZXB_EMRI_bumpy}.

To achieve these scientific goals, precise parameter extraction is essential. However, EMRI parameter estimation remains highly challenging~\cite{levati2026parameterestimationbiasinducedtransient,Speri_2026,cole2026sequentialsimulationbasedinferenceextreme}. The full parameter space is vast, while the true signal occupies only a tiny fraction of it, making the search intrinsically difficult~\cite{2026_EMRI_PE_intrinic}. Moreover, the likelihood surface is highly complex and features numerous local maxima arising from strong parameter degeneracies, where different parameter combinations can yield similarly good matches to the data. These secondary peaks may lie far from the true global maximum, complicating reliable signal identification and parameter recovery. As a result, the development of efficient and robust EMRI data-analysis methods remains a central challenge for future space-based gravitational-wave observations.

Previous works have adopted several strategies to address these challenges~\cite{2009CQGra..26m5004B,cole2026sequentialsimulationbasedinferenceextreme,2012_EMRI_F_statistic,2024_LB_EMRI_PE,2025_YQY_EMRI_PE,Zhao_2023,2025_ZXB_EMRI_bumpy,zou2026PSO}.
Ref.~\cite{2012_EMRI_F_statistic} used the phenomenological family of waveforms for detecting EMRI signals using the cumulative F-statistic.
Ref.~\cite{2024_LB_EMRI_PE} explored a machine learning approach to Bayesian posterior estimation of EMRI signals, leveraging the flow matching technique based on ordinary differential equation neural networks. 
Ref.~\cite{2025_YQY_EMRI_PE} used a convolutional neural network (CNN) to rapidly infer three key intrinsic parameters, including the mass and spin of the SMBH, as well as the initial orbital eccentricity.
Ref.~\cite{2026_EMRI_PE_intrinic} further introduced a hierarchical search framework capable of detecting EMRI signals in noisy data over wide prior ranges, successfully estimating 11 parameters for a non-spinning EMRI system with a signal-to-noise ratio (SNR) of 56.

In this work, based on deep learning, we successfully recover the 7 intrinsic parameters for EMRI systems involving spinning black holes without relying on detailed priors for extrinsic parameters. The overall parameter estimation algorithm consists of three stages.
 Since EMRI observations are heavily contaminated by detector noise, we first extract signal information from noisy short-time Fourier transform (STFT)~\cite{1171703,1162950,ozhan2022} images in the time-frequency domain using a U-Net architecture.
 Second, these features are then fed into a multi-channel convolutional neural network~\cite{Lecun1998, AlexNet2012,googleNet2014,resNet2015,VGG2015} (CNN) equipped with a dual attention mechanism (frequency attention and time attention) to recover the 7 intrinsic parameters. These parameters primarily determine the morphology of the time-frequency tracks, and the CNN-based estimation effectively narrows the search region, achieving a reduction of the 7-dimensional parameter space by a factor of $\sim 10^{4}$. 
  Finally, based on the reduced parameter-space ranges and frequency band, we repeat the CNN-based estimation procedure with a smaller search region: the ranges of $\{\log M, a, \mu, t_p, e_0, p_0 \}$ and ${x_I}_0$ are further reduced  by factors of $\sim 10^{2}$ and $\sim 10^{1}$, respectively, resulting in an overall reduction of the 7-parameter space  by a factor of $\sim 10^{8}$. By contrast, the remaining extrinsic parameters mainly affect the amplitudes of the STFT maps. We will recover these extrinsic parameters in future work.

This paper is organized as follows. Sec.~\ref{sec:EMRI simulation} introduces the data simulation of EMRIs. Sec.~\ref{sec:U-net_denoising} details the extraction of time-frequency tracks from noisy EMRIs using a U-Net framework. Sec.~\ref{sec:intrinsic_PE} describes the hierarchical parameter estimation strategy for intrinsic parameters based on a multi-channel CNN regression framework. Finally, Sec.~\ref{sec:conclusion} summarizes our results and outlines directions for future work.

\section{EMRI Data simulation}\label{sec:EMRI simulation}


For ground-based laser interferometers, laser frequency noise can be eliminated with extremely high precision by keeping the arm length stable at the picometer level. For space-borne GW detectors, however, due to the unequal-arm-length coupling effect caused by the relative motion of satellites, time-delay interferometry (TDI)~\cite{TDI2004,TDI2023,TDI2000,TDI1999} is required to achieve the target sensitivity. As a data post-processing technique, TDI constructs virtual equal-arm interferometers by applying time delays and linear combinations to the data streams.

\begin{figure}[H] 
\centering
    \includegraphics[width=0.45\textwidth]{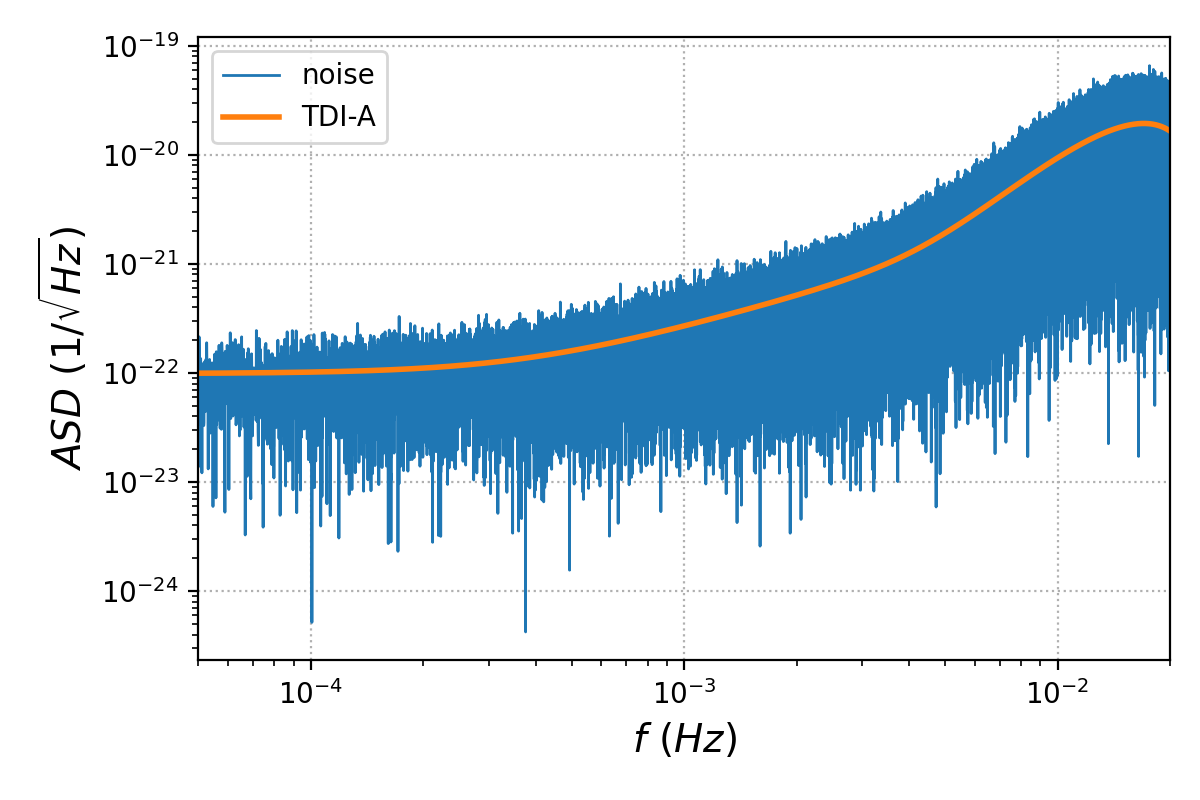}
    \caption{Comparison between the simulated noise and the  target noise ASD for the Taiji TDI-A channel. The simulated noise is shown in blue, while the orange curve denotes the target ASD derived from the Taiji instrument noise model~\cite{2026TaijiChallenge_DMH}.}
    \label{fig:noise}
\end{figure}

In general, the observed signal, $d(t)$, can be described as the sum of the GW signal $h(t)$ and the detector noise $n(t)$. 
\begin{equation}
    d_i(t)=h_i(t)+n_i(t),i=\{A,E,T\}
\end{equation}
where the value of $i$ represents different the second generation TDI channels $(A,E,T)$. In this work, we use the FastEMRIWaveform package $\textbf{FEW}$ ~\cite{FEW2025,2021_FEW1,2021_FEW2,2023_FEW} (5PN-AAK template) to obtain the EMRI waveform, which is described by 14 parameters:
\begin{equation}
\theta = \{ M,a,\mu, t_p, e_0,{x_I}_0 , \Phi_{r_0},\Phi_{\theta_0},\Phi_{\phi_0},\theta_K,\phi_K,\theta_S, \phi_S, D_L\}.
\end{equation}
Here, $M$ and $a$ represent the mass and spin of the central supermassive black hole, while $\mu$ denotes the mass of the secondary object. The orbital elements are characterized by the time to plunge $t_p$, initial eccentricity $e_0$, and the cosine of the initial inclination of the orbital plane ${x_I}_0  = \cos \iota_0$. The initial semi-latus rectum $p_0$ is then determined numerically from the set of \{$t_{p}$, $\mu$, $M$, $e_0$, $a$, ${x_I}_0 $\}~\cite{2026_EMRI_PE_intrinic}. The initial radial, polar, and azimuthal orbital phases are $\Phi_{r_0}$, $\Phi_{\theta_0}$, and $\Phi_{\phi_0}$. The orientation of the orbital angular momentum is specified by $\theta_K$ and $\phi_K$. The source's sky location in the Solar System barycenter frame is given by the angles $\theta_S$ and $\phi_S$. The luminosity distance is denoted by $D_L$. 

We use the \textbf{Triangle-Simulator}~\cite{2026TaijiChallenge_DMH} package to incorporate the Taiji response and generate the TDI outputs, producing both the signal $h_i(t)$ and the instrument noise $n_i(t)$. In this work, we focus on the TDI $A$ and $E$ channels, which are sensitive to gravitational waves and share similar amplitude spectral densities (ASDs), while the $T$ channel is insensitive to gravitational waves and is used to characterize the instrument noise. 
Fig.~\ref{fig:noise} presents the spectrum of the simulated noise (blue) for the Taiji TDI-A channel alongside the theoretical target (orange). 


\section{Signal extraction via U‑Net network} \label{sec:U-net_denoising} 

 To train the model, we first construct an EMRI dataset (see Sec.~\ref{sec:EMRI simulation}) comprising paired samples of noisy spectrograms and their corresponding clean signal masks. The parameter ranges used for data generation are summarized in Tab.~\ref{tab:injection}. In total, we generate 20,000 paired samples with SNRs ranging from 50 to 100. The dataset is randomly partitioned into training, validation, and test sets at a ratio of 70\%, 15\%, and 15\%, respectively.

Effective noise reduction is essential for the accurate detection and analysis of EMRI signals, particularly when they are deeply buried in detector noise. In this work, we employ a denoising approach 
using a U-Net~\cite{Unet2015} architecture. Unlike conventional filtering methods, which often leave residual noise, the U-Net learns a direct mapping from noisy time-frequency representations (e.g., spectrograms) to clean signal structures.

\begin{table}[H]
\footnotesize
\begin{threeparttable}\caption{The parameter ranges for EMRI dataset.}\label{tab:injection}
\doublerulesep 0.1pt \tabcolsep 25pt 
\begin{tabular}{llllll}
\toprule
           Parameter &  Range  \\ \hline
  $\mu$ ($\textup{M}_\odot$) & [20,100]\\
  $M$ ($\textup{M}_\odot$) & [$10^{6.0}$,$10^{7.0}$] \\
  $e_0$ &  [0.1,0.6]\\
  $a$ &  [0.1,0.9]\\
  ${x_I}_0 $ &  [0.1,1.0]\\
  $t_{p}$ (yr) &  [0.4,0.46]\\
  $p_{0}$ &  set by \{$t_{p}$, $\mu$, $M$, $e_0$, $a$, ${x_I}_0 $\} \\
        $\Phi_{r_0}$&          [0,$2\pi$]\\
        $\Phi_{\theta_0}$&          [0,$2\pi$]\\
        $\Phi_{\phi_0}$&          [0,$2\pi$]\\
    $\theta_K$  &    [0,$\pi$]\\
  $\phi_K$&      [0,$2\pi$]\\
  $ \theta_S$  &     [0,$\pi$]\\
  $\phi_S$&      [0,$2\pi$]\\
  $\theta_{\rm{lat}} = \pi/2 - \theta_S$  & [$-\pi/2$,$\pi/2$] \\
  $\phi_{\rm{lon}} = \phi_S$  &  [0,$2\pi$] \\
  SNR &  [50,100] \\
  $D_L$ (Gpc) & set by SNR \\
\bottomrule
\end{tabular}
\end{threeparttable}
\end{table}

\begin{figure}[H] 
\centering
	\includegraphics[width=0.45\textwidth]{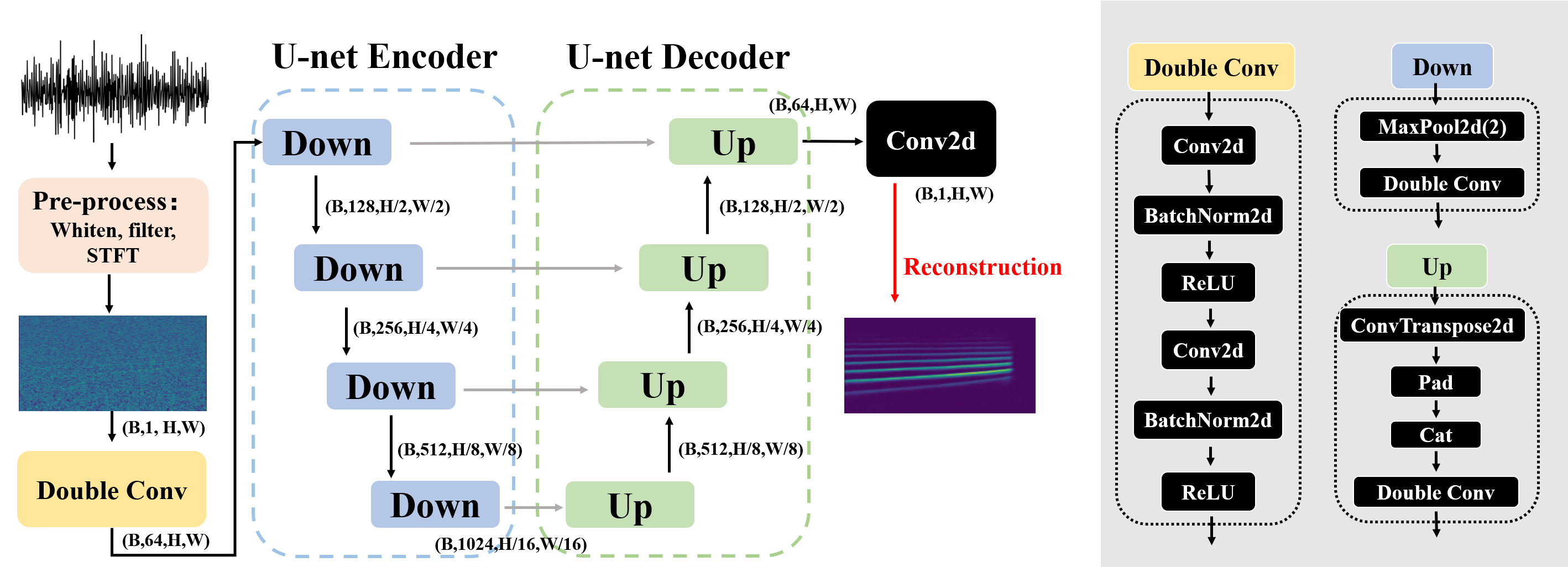}
    \caption{U-Net architecture adopted in this work. The code is available at \url{https://github.com/milesial/Pytorch-UNet}~\cite{Unet2015}.}
    \label{fig:U-net}
\end{figure}

Fig.~\ref{fig:U-net} illustrates the overall workflow of our signal extraction method, encompassing data pre-processing, encoding, decoding, and output generation. For clarity, the figure depicts the pipeline for a single spectrogram; in practice, the four multi-scale spectrograms are stacked along the channel dimension and processed in parallel by the same network.

(1) Pre-processing: To enhance the signal components and suppress the colored noise background, we first apply whitening and filtering, and then generate the STFT images. The time step is \(dt = 25~\mathrm{s}\), and the STFT window lengths are \(\{2048, 4096, 6144, 8192\}\) samples, corresponding to \(T_{\mathrm{window}} \approx 0.59\)–\(2.37~\mathrm{day}\). A larger window gives higher frequency resolution but lower time resolution, and vice versa.
The resulting four multi-scale STFT spectrograms are then fed into the U-Net architecture for signal extraction.

(2) Encoding: The encoder path first applies a \textbf{Double Conv} module to project the single-channel input spectrogram to 64 feature channels, and then consists of four sequential \textbf{Down} blocks. Each \textbf{Down} block first applies a 2×2 max-pooling operation to halve the spatial dimensions, followed by a \textbf{Double Conv} module (two consecutive 3×3 convolutions, each followed by batch normalization and ReLU activation). This structure progressively reduces the feature map size from (B, 64, H, W) to (B, 1024, H/16, W/16), where B denotes the batch size, H and W represent the height and width of the input spectrogram, and the second dimension (64 and 1024) indicates the number of feature channels. The channel depth doubles at each stage, enabling the network to learn increasingly abstract representations of the signal.

(3) Decoding: The decoder path symmetrically restores the spatial resolution using four \textbf{Up} blocks. Each Up block begins with a transposed convolution (\textbf{ConvTranspose2d}) to upsample the feature map by a factor of two, followed by padding and concatenation (Cat) with the corresponding high-resolution feature map from the encoder via skip connections. The fused features are then processed by a \textbf{Double Conv} module to refine the spatial details. This design ensures that fine-grained time-frequency structures are preserved during reconstruction.

(4) Output generation: A 1×1 2D convolution projects the 64-channel decoder features to the target single-channel space, yielding an output of shape (B, 1, H, W) that represents the reconstructed spectrogram. The network is trained to minimize the MSE between the predicted and clean spectrograms, enabling it to recover the signal content while attenuating the noise at inference time.

\begin{figure}[H] 
\centering
	\includegraphics[width=0.48\textwidth]{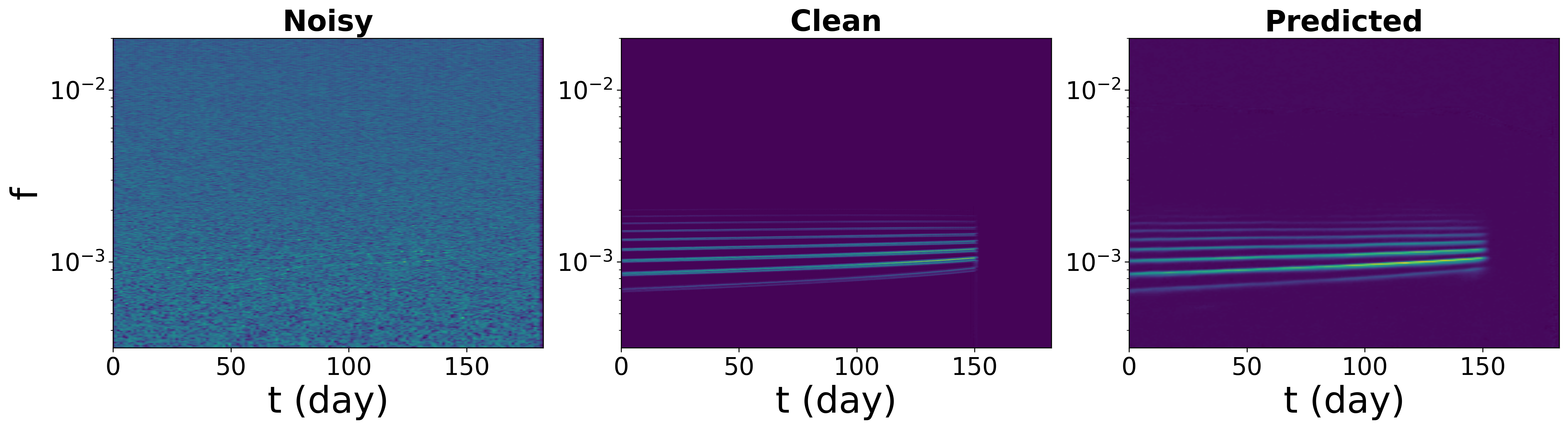}
    \caption{Comparison between the clean/noisy EMRI signal and the U-Net predicted one in the time-frequency domain. The first column shows the noisy input data, the second column shows the injected clean signal, and the third column presents the signal reconstructed by the U-Net model. The combined SNR of TDI-A and TDI-E channels is 54. Parameters of the injected EMRI are set by
    $t_p=0.414127492~\rm{yr}$, $\log M=6.592691615$, $\mu=36.16054057~M_\odot$, $a=0.246946585$, $e_0 = 0.328771766$, ${x_I}_0  =0.296935802$, $\Phi_{r_0}=0.785985654 $, $\Phi_{\theta_0}=4.476201891$, $\Phi_{\phi_0}= 3.868663471$, $\theta_K=1.752404985$, $\phi_K=4.459292421$, $\theta_{\rm{lat}}=1.520039887$, $\phi_{\rm{lon}}=0.679193971$, and $D_L=0.66129397~\rm{Gpc}$. The time step is \(dt = 25~\mathrm{s}\), and the STFT window length is \(8192\) samples, corresponding to \(T_{\mathrm{window}} = 8192 \times 25~\mathrm{s} \approx 2.37~\mathrm{day}\).
} 
    \label{fig:Unet_extract}
\end{figure}

Fig.~\ref{fig:Unet_extract} compares the clean, noisy, and U-Net reconstructed signals in the time-frequency domain. From left to right, the three columns correspond to the noisy input, the injected clean signal, and the U-Net output, respectively. 
It is clearly observed that the U-Net effectively recovers the signal with high fidelity under the high-SNR condition ($\rm{SNR}>50$).

To quantify the extraction capability of the U-Net for EMRI signals under varying SNR conditions, we construct a new dataset comprising 9,000 EMRI samples with the combined SNR of TDI-A and TDI-E channels ranging from 10 to 100. We evaluate the similarity between the injected clean signal and the U-Net-extracted one using cosine similarity, which is computed on the flattened spectrograms:

\begin{equation}
S_{\rm cos} = \frac{\mathbf{s}_{\text{inj}} \cdot \mathbf{s}_{\text{ext}}}{\|\mathbf{s}_{\text{inj}}\| \|\mathbf{s}_{\text{ext}}\|} = \frac{\sum_{i=1}^{N} s_{\text{inj}}^{(i)} \cdot s_{\text{ext}}^{(i)}}{\sqrt{\sum_{i=1}^{N} \left(s_{\text{inj}}^{(i)}\right)^2} \sqrt{\sum_{i=1}^{N} \left(s_{\text{ext}}^{(i)}\right)^2}}
\end{equation}
where $\mathbf{s}_{\text{inj}}$ and $\mathbf{s}_{\text{ext}}$ are the vectorized spectrograms of the injected and extracted signals, respectively, and $N$ is the number of elements in the vector. Cosine similarity values approaching $1$ indicate higher consistency between the reconstructed and injected signals, demonstrating the effectiveness of the U-Net model in EMRI signal recovery. Since the signal occupies only a small region of the spectrogram, we compute the cosine similarity exclusively over the largest 30\% of the spectral area.

\begin{figure}[H] 
\centering
	\includegraphics[width=0.45\textwidth]{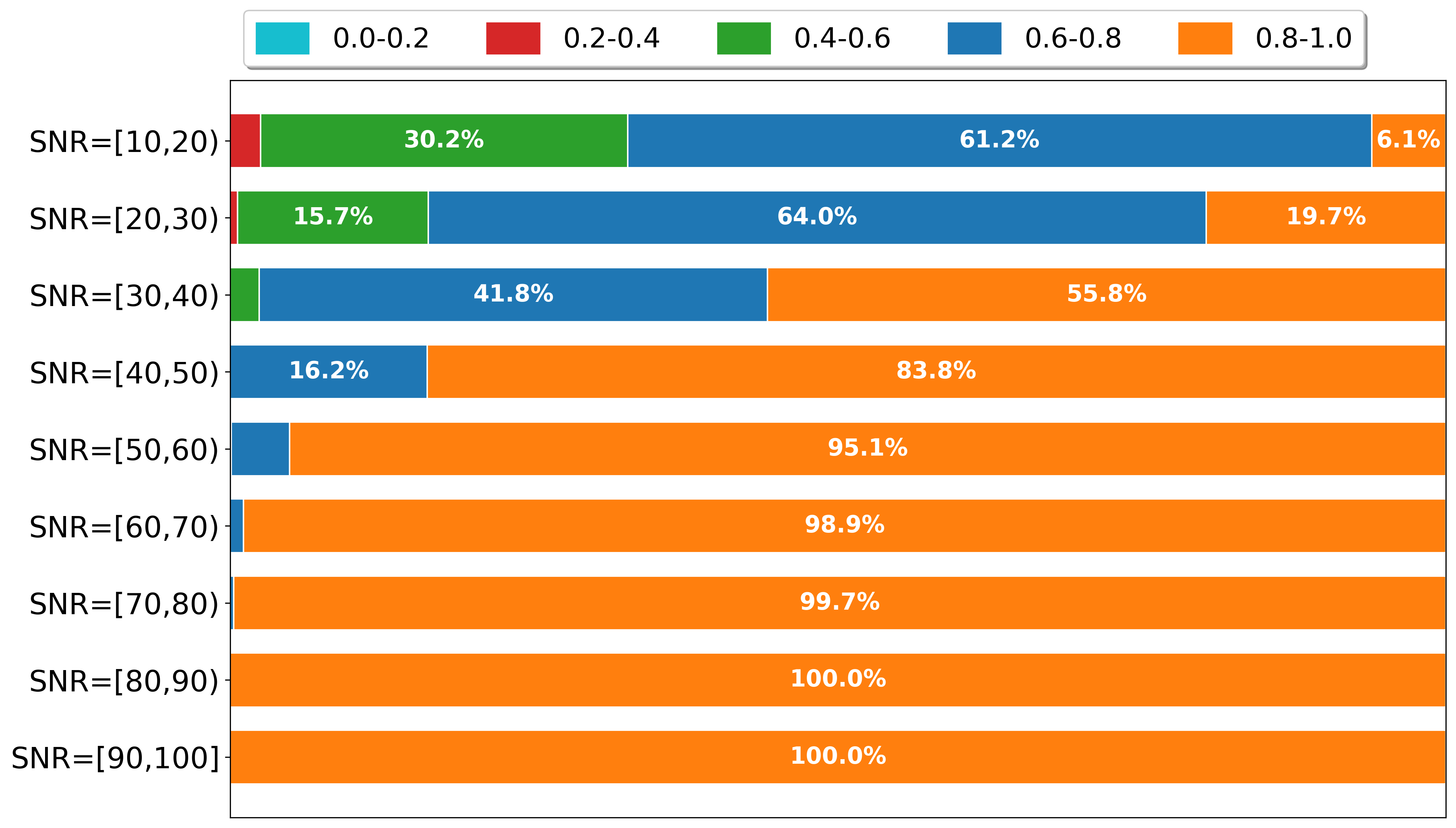}
    \caption{Cosine similarity between the injected clean signal and the one extracted by U-Net. The color bar indicates the cosine similarity range, where values closer to 1 correspond to higher similarity between the two signals.  For illustration, we present results exclusively for the TDI-A channel data, with the STFT window length set to 8192 samples and the time step \(dt = 25~\mathrm{s}\).}
    \label{fig:Scos}
\end{figure}

We adopt a cosine similarity threshold of 0.8 to define successful reconstruction. As shown in Fig.~\ref{fig:Scos}, the success rate exceeds 80\% for signals with $\rm{SNR} > 40$, and surpasses 95\% for golden EMRI events ($\rm{SNR} > 50$). However, performance drops sharply when SNR falls below 40. These results demonstrate that the U-Net effectively recovers golden EMRI signals in the time-frequency domain, preserving track continuity and fine-grained features while suppressing background noise. Overall, the U-Net provides a robust and accurate framework for EMRI signal denoising under moderate-to-high SNR conditions.

\section{Intrinsic parameters estimation } \label{sec:intrinsic_PE}


\begin{figure}[H]
\centering
\begin{minipage}{0.8\linewidth}
\centering
\includegraphics[width=\linewidth,height=0.3\textwidth]{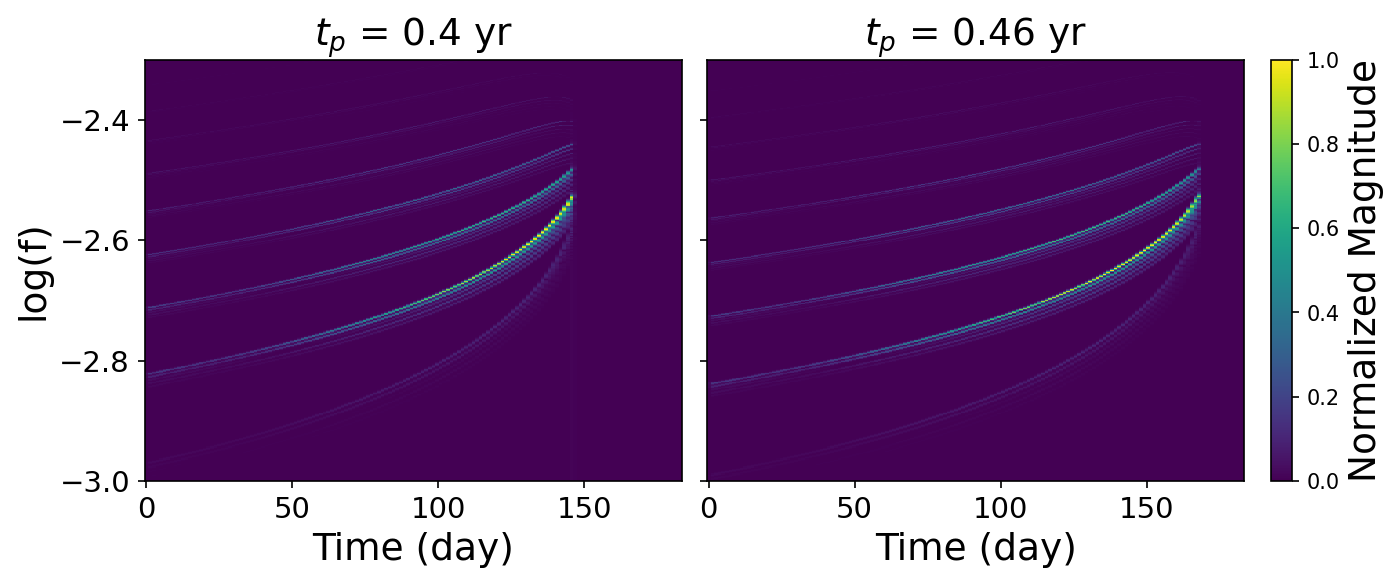}\\
(a)
\end{minipage}
\begin{minipage}{0.8\linewidth}
\centering
\includegraphics[width=\linewidth,height=0.3\textwidth]{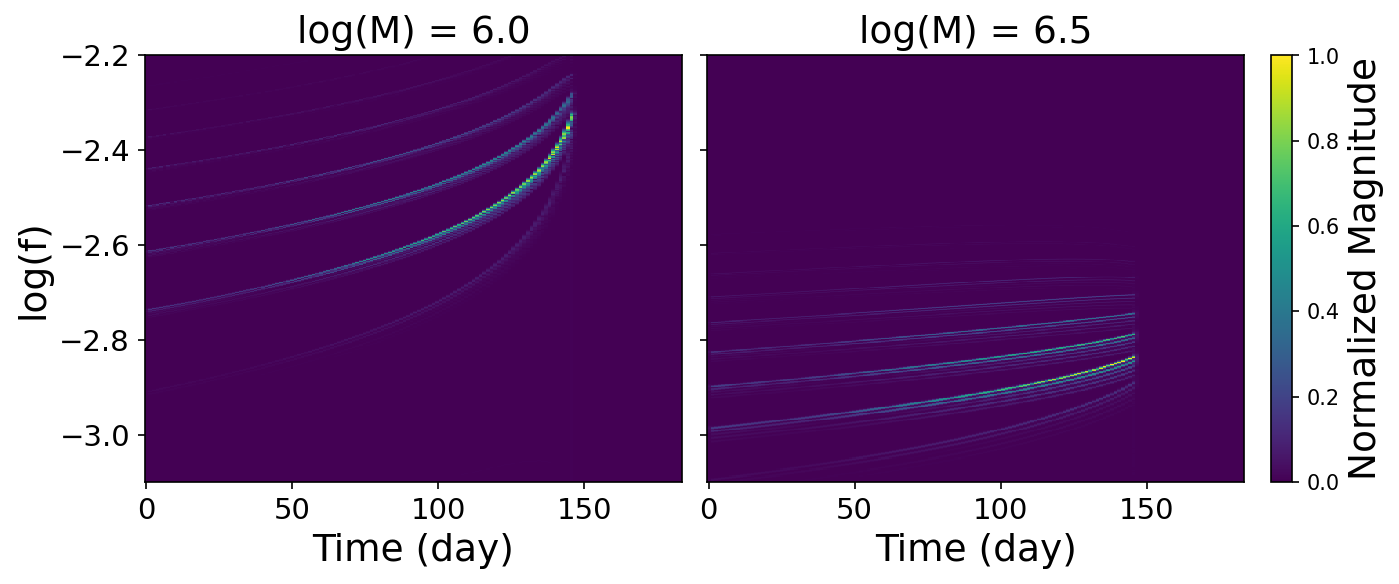}\\
(b)
\end{minipage}
\begin{minipage}{0.8\linewidth}
\centering
\includegraphics[width=\linewidth,height=0.3\textwidth]{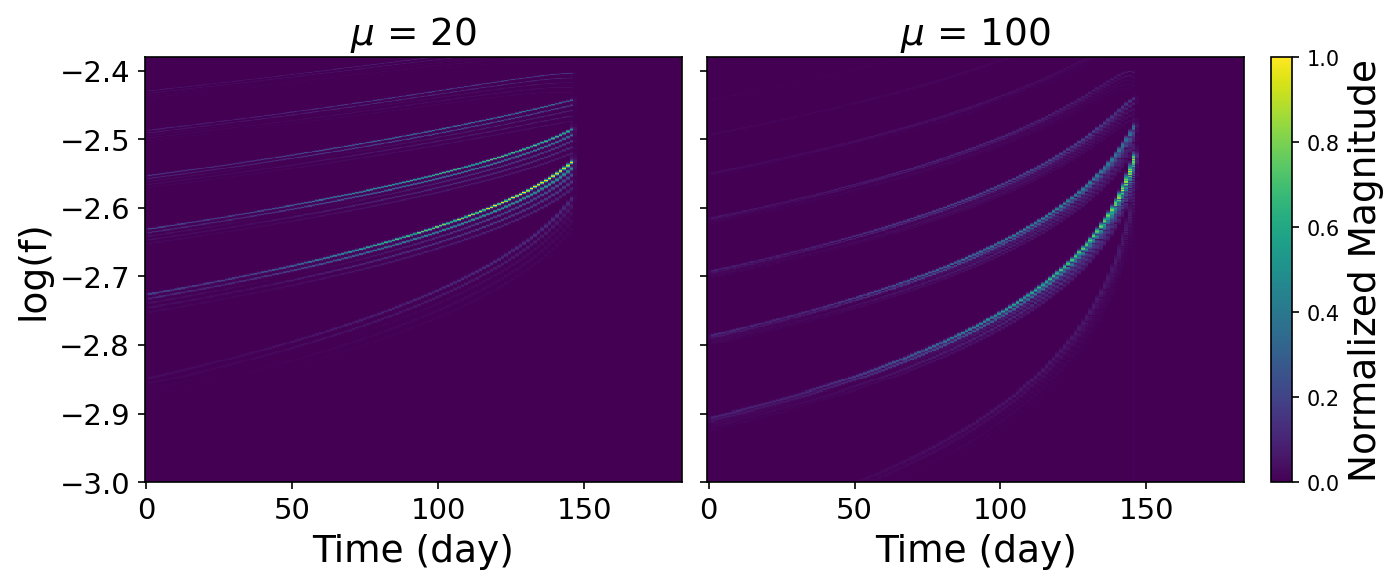}\\
(c)
\end{minipage}
\begin{minipage}{0.8\linewidth}
\centering
\includegraphics[width=\linewidth,height=0.3\textwidth]{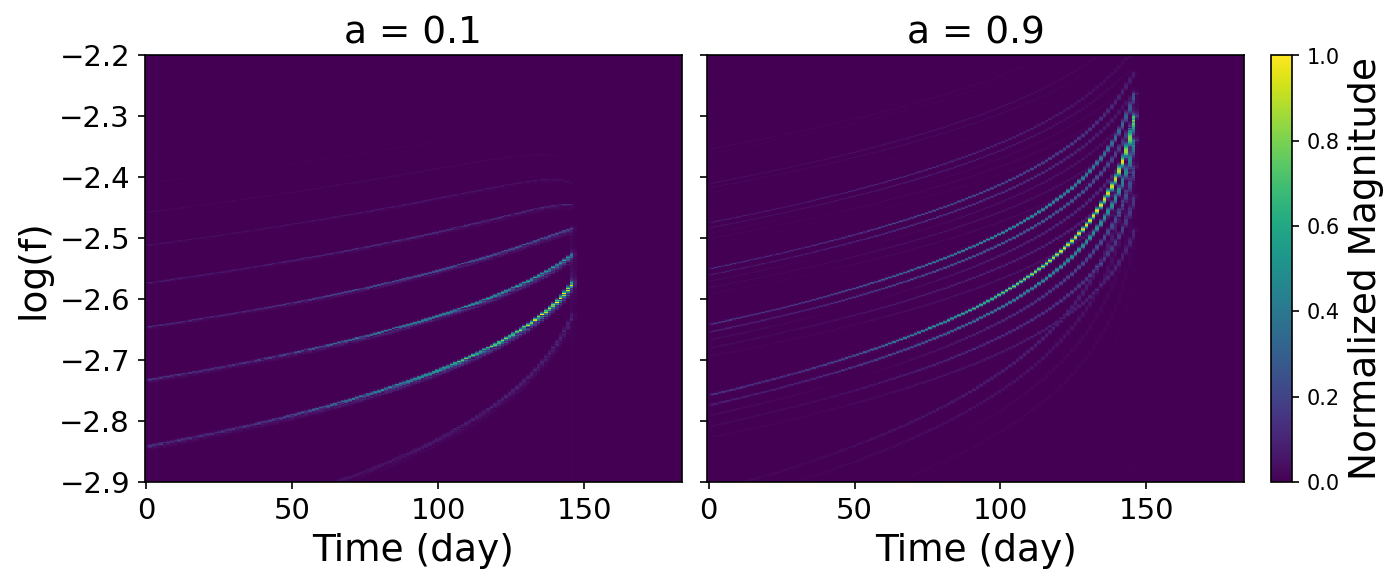}\\
(d)
\end{minipage}
\begin{minipage}{0.8\linewidth}
\centering
\includegraphics[width=\linewidth,height=0.3\textwidth]{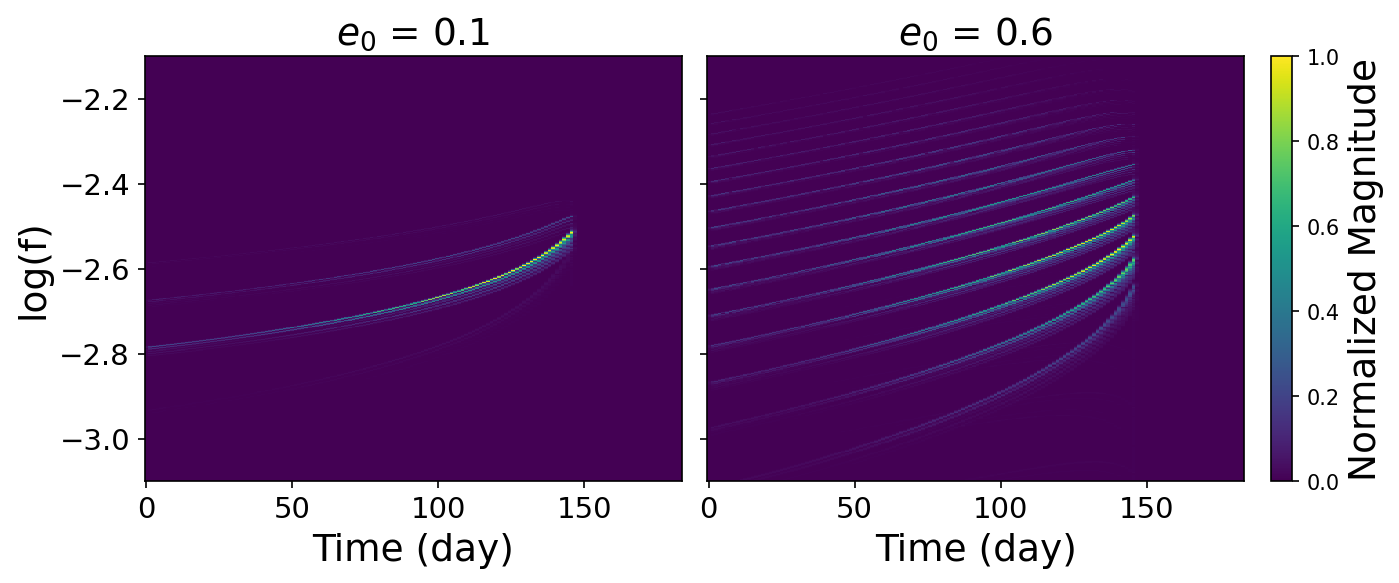}\\
(e)
\end{minipage}
\begin{minipage}{0.8\linewidth}
\centering
\includegraphics[width=\linewidth,height=0.3\textwidth]{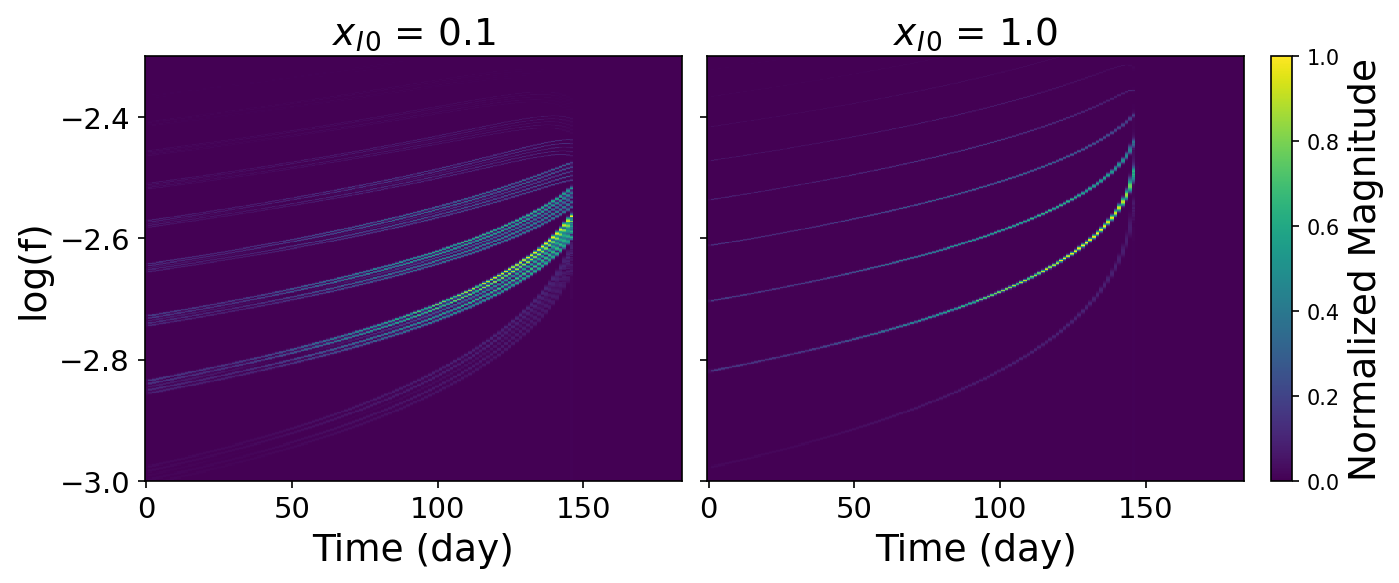}\\
(f)
\end{minipage}
\caption{Sensitivity of STFT spectrograms to intrinsic parameters of EMRIs.  
The extrinsic parameters are fixed as $\Phi_{r_0},\Phi_{\theta_0},\Phi_{\phi_0}= 0.6, 1.1, 2.3$, $\theta_K,\phi_K,\theta_{\rm{lat}},\phi_{\rm{lon}}=0.5, 0.7, 0.6, 0.4$, $D_L=1.0~\rm{Gpc}$.  
(a) $t_p=\{0.4,0.46\}~\rm{yr}$, $M=10^{6.2}~M_\odot$, $\mu=50~M_\odot$, $a=0.3$, $e_0 = 0.3$, ${x_I}_0  = 0.5$.  
(b) $t_p=0.4~\rm{yr}$, $\log M=\{6.0,6.5\}$, $\mu=50~M_\odot$, $a=0.3$, $e_0 = 0.3$, ${x_I}_0  = 0.5$.  
(c) $t_p=0.4~\rm{yr}$, $\log M=6.2$, $\mu=\{20,100\}~M_\odot$, $a=0.3$, $e_0 = 0.3$, ${x_I}_0  = 0.5$.  
(d) $t_p=0.4~\rm{yr}$, $\log M=6.2$, $\mu=50~M_\odot$, $a=\{0.1,0.9\}$, $e_0 = 0.3$, ${x_I}_0  = 0.5$.  
(e) $t_p=0.4~\rm{yr}$, $\log M=6.2$, $\mu=50~M_\odot$, $a=0.3$, $e_0 =\{0.1,0.6\}$, ${x_I}_0  = 0.5$.  
(f) $t_p=0.4~\rm{yr}$, $\log M=6.2$, $\mu=50~M_\odot$, $a=0.3$, $e_0 = 0.3$, ${x_I}_0 = \{0.1,1.0\}$. The initial semi-latus rectum $p_0$ is determined numerically from the set of $\{t_{p}$, $\mu$, $M$, $e_0$, $a$, ${x_I}_0 \}$ using \textbf{FEW} package~\cite{FEW2025,2026_EMRI_PE_intrinic}.
}
\label{fig:intrinic}
\end{figure}

In this section, we design a multi-channel CNN regression framework for intrinsic parameter estimation of EMRI systems based on the time–frequency images obtained by the U-Net model (Sec.~\ref{sec:U-net_denoising}).  We partition the parameter space into intrinsic parameters and  extrinsic parameters. The intrinsic parameters primarily determine the morphology and trajectory of the EMRI signal in the time–frequency domain, including the mass $M$ and spin $a$ of the central SMBH, the mass $\mu$ of the small compact object, the time to plunge $t_p$, the initial orbital parameters: eccentricity $e_0$, inclination angle $\iota_0$ (${x_I}_0=\cos{\iota_0}$), and  the semi-latus rectum $p_0$ (determined by \{$t_{p}$, $\mu$, $M$, $e_0$, $a$, ${x_I}_0 $\} ). In contrast, the extrinsic parameters mainly affect the amplitude of the spectrogram.

Fig.~\ref{fig:intrinic} presents a comparison of EMRI STFT spectrograms under variations of individual intrinsic parameters. Each of the 6 subplots illustrates the effect of varying one intrinsic parameter while keeping all other parameters fixed. 
The results show that different intrinsic parameters induce distinct and observable modifications in the EMRI signal tracks, which enable CNN to effectively learn discriminative features from the time–frequency representations and rapidly constrain the intrinsic parameter space.

\subsection{Step 1: Rough Estimation}\label{sec:step1}
We construct a new dataset for deep-learning-based intrinsic parameter estimation, comprising 80,000 samples. The dataset is randomly split into training, validation, and test sets with a ratio of 70\%, 15\%, and 15\%, respectively, corresponding to 56,000, 12,000, and 12,000 samples. The combined SNR of the TDI-A and TDI-E channels ranges from 50 to 100. 
For each TDI channel, we generate 4 STFT images with different window lengths, namely \(2048\), \(4096\), \(6144\), and \(8192\) samples. This yields 8 input images per sample in total.

\begin{figure}[H] 
\centering
	\includegraphics[width=0.48\textwidth]{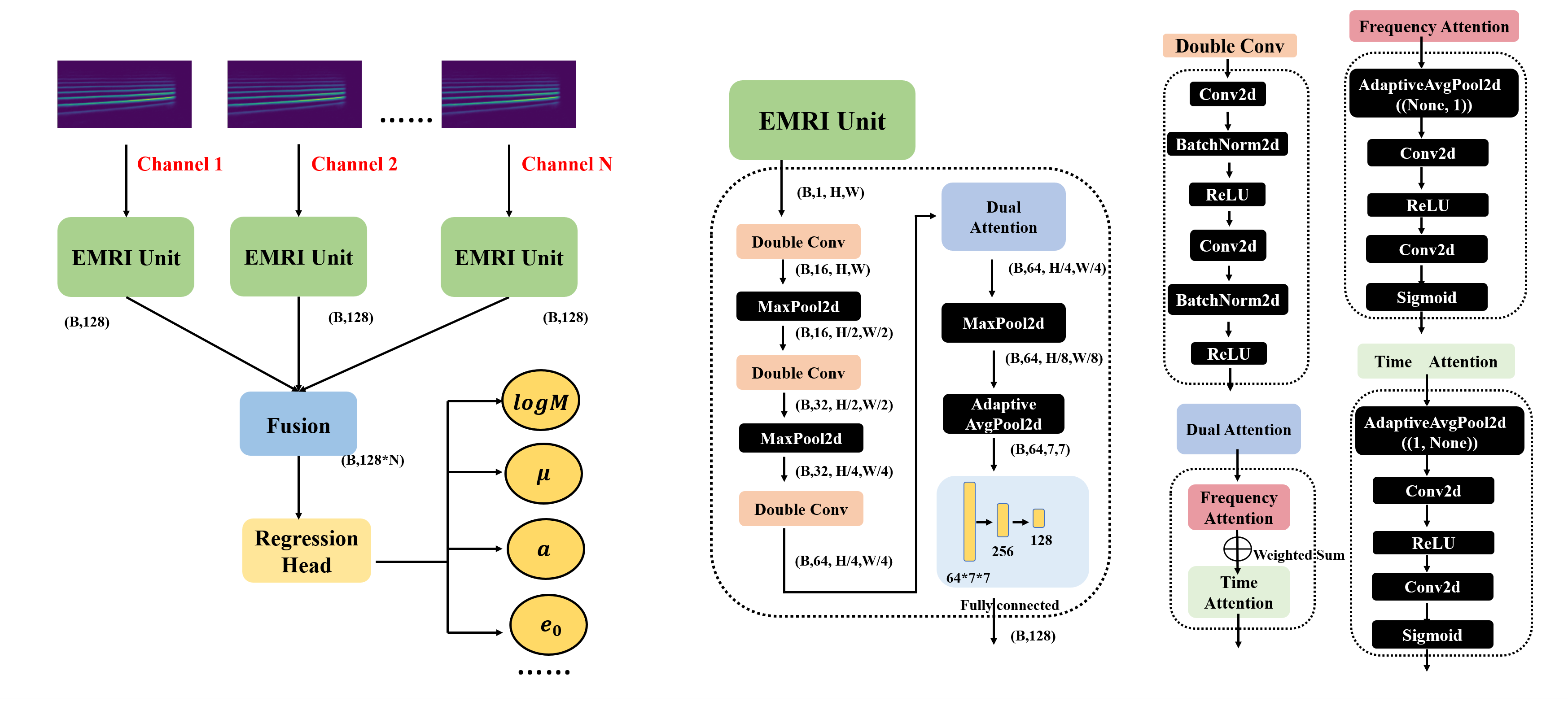}
    \caption{The multi-channel CNN regression framework for intrinsic parameter estimation. The network processes $N$ channels in parallel through the \textbf{EMRI Unit} and concatenates their extracted features via a \textbf{Fusion} layer. Each \textbf{EMRI Unit} incorporates a hierarchical \textbf{Double Conv} architecture and a \textbf{Dual  attention} mechanism to adaptively highlight significant time–frequency ridges. Finally, a \textbf{Regression Head} maps the fused high-dimensional representations to the physical source parameters.
    }
    \label{fig:our CNN}
\end{figure}

Fig.~\ref{fig:our CNN} illustrates the proposed CNN framework designed for estimating the intrinsic parameters of EMRIs. The network operates by performing regression on time–frequency representations of the EMRI signals. Specifically, each input channel is processed independently by a dedicated convolutional encoder, termed the \textbf{EMRI Unit}. The high-level features extracted from all channels are subsequently concatenated in a \textbf{fusion} layer and passed to a \textbf{regression head}, which outputs the estimated source parameters (e.g.\{$t_p$, $\mu$, $\log M$, $e_0$, $a$, $p_0 $\}).

\begin{figure}[H] 
\centering
	\includegraphics[width=0.46\textwidth]{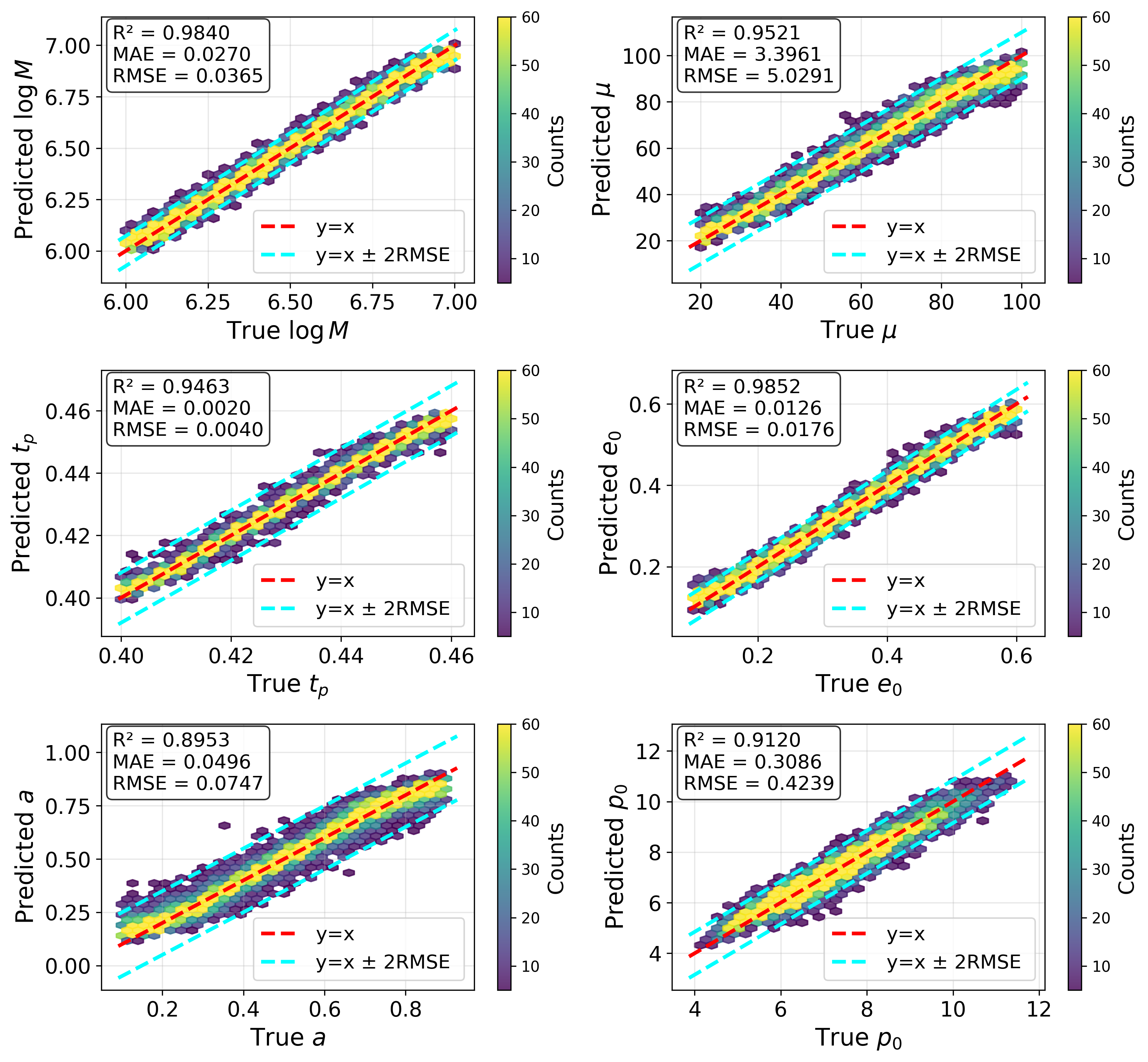}
    \caption{Recovery performance of the multi-channel CNN regression framework for the intrinsic parameters $t_p$, $\log M$, $\mu$, $a$, $e_0$, and ${p}_0$ (Step 1). The predicted values (y-axis) are plotted against the corresponding true values (x-axis). The red dashed line denotes perfect agreement ($y=x$), and the cyan dashed lines indicate the $y=x\  \pm \ 2\ \text{RMSE}$ bounds. Notably, 95.1\% of the data points fall within the $2\,\text{RMSE}$ bounds. The color bar shows the number of events in each hexagonal bin. }
    \label{fig:intri_result}
\end{figure}

The architecture of the \textbf{EMRI Unit} follows a hierarchical encoder design composed of multiple stages.
Each stage consists of a \textbf{Double Conv} block comprising two consecutive 3×3 convolutional layers, each followed by batch normalization and ReLU activation to extract local features. These blocks are interleaved with 2×2 max-pooling operations to progressively downsample the feature maps and expand the receptive field. 
This structure enables the network to capture multi-scale representations from the input spectrograms effectively.

To further enhance feature discriminability, the unit incorporates a \textbf{Dual Attention} mechanism that explicitly decouples the temporal and spectral dimensions via orthogonal pooling strategies. Specifically, the \textbf{frequency-aware} branch compresses the temporal dimension to aggregate global temporal context and generate attention maps along the frequency axis. Conversely, the \textbf{time-aware} branch compresses the spectral dimension to capture long-range spectral dependencies and produce attention maps along the time axis. Each branch refines its features through two successive 1×1 convolutions separated by a single ReLU activation, followed by a Sigmoid gating function. The two resulting attention maps are then fused with equal weights ($\alpha=0.5$) and applied to re-weight the input features via element-wise multiplication. Following the attention module, the extracted feature maps undergo MaxPool2d downsampling and AdaptiveAvgPool2d operations. The pooled features are then flattened and passed through a fully connected regression head, progressively compressed from 256 to 128 dimensions, to ultimately output the 128-dimensional latent representations for parameter estimation.

To quantitatively assess the reconstruction performance, we adopt the following metrics.\\ (1) The coefficient of determination ($R^2$) measures the proportion of variance in the predicted values explained by the true values, defined as:
\begin{equation}
R^2 = 1 - \frac{\sum_{i=1}^{N} (y_i - \hat{y}_i)^2}{\sum_{i=1}^{N} (y_i - \bar{y})^2}
\end{equation}
where $y_i$   and $\hat{y}_i$   are the true and predicted values, respectively. 
$\bar{y}$ is the mean of the true values, and 
$N$ is the number of samples. The coefficient of $R^2$ measures the improvement in predictive performance of the model relative to simply using the mean as a predictor: \( R^2 = 1 \) indicates a perfect fit; \( R^2 = 0 \)  implies performance equivalent to the mean predictor; and \( R^2 < 0 \) suggests the model performs worse than the mean baseline. Therefore, the closer \( R^2 \) is to 1, the more significantly the model captures parameter variations beyond a naive mean estimate.\\
(2) The mean absolute error (MAE) quantifies the average absolute deviation between predictions and true values:
\begin{equation}
\text{MAE} = \frac{1}{N} \sum_{i=1}^{N} |y_i - \hat{y}_i|
\end{equation}
 It imposes a linear penalty on errors, treating all deviations equally, which makes it robust to outliers and easily interpretable in the original data units.\\
(3) The root mean square error (RMSE) is defined as:
\begin{equation}
\text{RMSE} = \sqrt{\frac{1}{N} \sum_{i=1}^{N} (y_i - \hat{y}_i)^2}
\end{equation}
RMSE gives greater weight to large errors due to the squaring operation, making it sensitive to outliers and particularly useful when large prediction errors are undesirable.

These metrics are reported in the upper-left corner of each subplot in Fig.~\ref{fig:intri_result}, which presents the predictions of our multi-channel CNN regression  model for the 6 EMRI parameters ( $\log M$, $\mu$, $t_p$,  $e_0$, $a$, and $p_0$) against their true values. 
For the 5 parameters ($\log M$, $\mu$, $t_p$, $e_0$, and $p_0$), the high-density hexagonal bins (yellow) align closely with the diagonal, corresponding to $R^2 > 0.9$, which indicates high-precision predictions for the majority of events. Furthermore, 95.1\% of the data points fall within the $2\,\text{RMSE}$ bounds, demonstrating the robustness of the model. 
In contrast, the reconstruction performance for the parameter $a$ is suboptimal; this will be addressed and improved in Step 2 (see Sec.~\ref{sec:step2}), alongside $x_{I0}$.

\begin{table}[H]
\footnotesize
\begin{threeparttable}\caption{The injected and recovered parameters of the EMRI signal shown in Fig.~\ref{fig:Unet_extract}. The combined SNR of the TDI-A and TDI-E channels is 54. The relative error $\mathcal{E}$ is defined as $\mathcal{E} = |P_{\rm inj}-P_{\rm est}|/P_{\rm inj}$.}\label{tab:intrinsic_result}
\doublerulesep 0.1pt \tabcolsep 4.5pt 
\begin{tabular}{lccccc}
\toprule
Param. & Prior & $P_{\rm inj}$ & $P_{\rm est}$ & RMSE & $\mathcal{E}$ \\ \hline
$\log M$   & [6.0,7.0] & 6.592692  & 6.6083603  & 0.0377 &  0.238\%   \\
$\mu$      & [20,100]  & 36.160541  & 38.299454  & 5.0599 &  5.915\%   \\
$t_{p}$    & [0.40,0.46]& 0.414127  & 0.41448227 & 0.0041 &  0.0857\%  \\
$e_0$      & [0.1,0.6] & 0.328772 & 0.32169452 & 0.0186 &  2.153\%   \\
$a$        & [0.1,0.9] & 0.246946  & 0.24807957 & 0.0750  &  0.459\%   \\
$p_0$        & [3.8,11.7] & 6.910498  & 6.735895 & 0.4239&  2.527\%   \\
\bottomrule
\end{tabular}
\end{threeparttable}
\end{table}

\begin{figure}[H] 
\centering
	\includegraphics[width=0.48\textwidth]{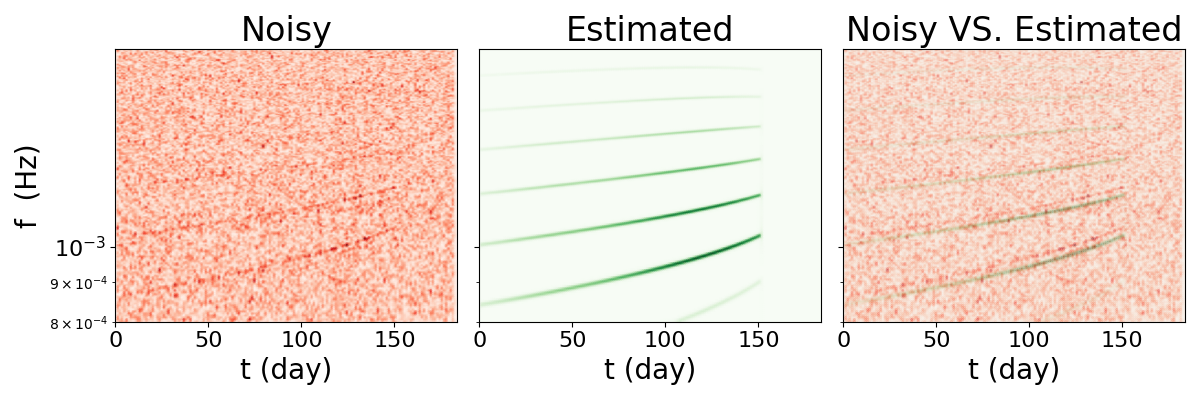}
    \caption{Time-frequency domain comparison between the noisy EMRI signal and the CNN-reconstructed signal (Step 1). 
The intrinsic parameters ($\log M$, $\mu$, $t_p$, $e_0$, and $a$) of the recovered waveform are assigned the CNN-predicted values $P_{\rm est}$ from Step 1, with $D_L = 1\,\text{Gpc}$. The remaining parameters are set to the median of their prior distributions.
} 
    \label{fig:our CNN_result_compare}
\end{figure}

Specifically, for the EMRI signal shown in Fig.~\ref{fig:Unet_extract}, Tab.~\ref{tab:intrinsic_result} lists the prior ranges, injected true values ($P_{\rm inj}$), estimated values ($P_{\rm est}$), RMSE , and the relative error $\mathcal{E}$ for 6 parameters. 
Fig.~\ref{fig:our CNN_result_compare} compares the noisy EMRI signal and our CNN-estimated one in the time-frequency domain. Our CNN-estimated EMRI signal exhibits strong agreement with the noisy signal in the time-frequency domain, with the main frequency evolution being well captured. However, small discrepancies can still be observed between the estimated and noisy signals. These residual differences indicate that there is still room for improvement in the estimation accuracy. In the following (see Sec.~\ref{sec:step2}),we will perform additional training on a reduced range to further improve the estimation accuracy.

\subsection{Step 2: Precise Estimation }\label{sec:step2}

In Step 2, we refine the parameter estimation by leveraging the narrowed parameter space and frequency band identified in Step 1. 
We construct a new dataset comprising 40,000 samples for this purpose. 
The parameter ranges for $\{\log M, \mu, e_0, a\}$ are defined as $\mathcal{U}[P_{\rm est} - 2\,\text{RMSE}, P_{\rm est} + 2\,\text{RMSE}]$, where $P_{\rm est}$ denotes the initial predictions obtained in Step 1 (see Tab.~\ref{tab:intrinsic_result}). 
For the time to plunge $t_p$, we retain the original parameter range used in Step 1, since the frequency-band restriction does not directly improve time-resolution and a hard truncation of $t_p$  could bias the refinement.
The frequency band is restricted to $[6.5\times 10^{-4}, 2.1\times 10^{-3}]\,\text{Hz}$. 
The dataset is randomly partitioned into training, validation, and test sets with a ratio of 70\%, 15\%, and 15\%, respectively. 
The combined SNR of the TDI-A and TDI-E channels ranges from 50 to 100. 
For each TDI channel, we generate 4 STFT images using window lengths of 2048, 4096, 6144, and 8192 samples, yielding a total of 8 input images per sample. 
Following the same process as Step 1, we perform parameter estimation for 7 parameters (including $x_{I0}$) to achieve refined results.

\begin{figure}[H] 
\centering
	\includegraphics[width=0.46\textwidth]{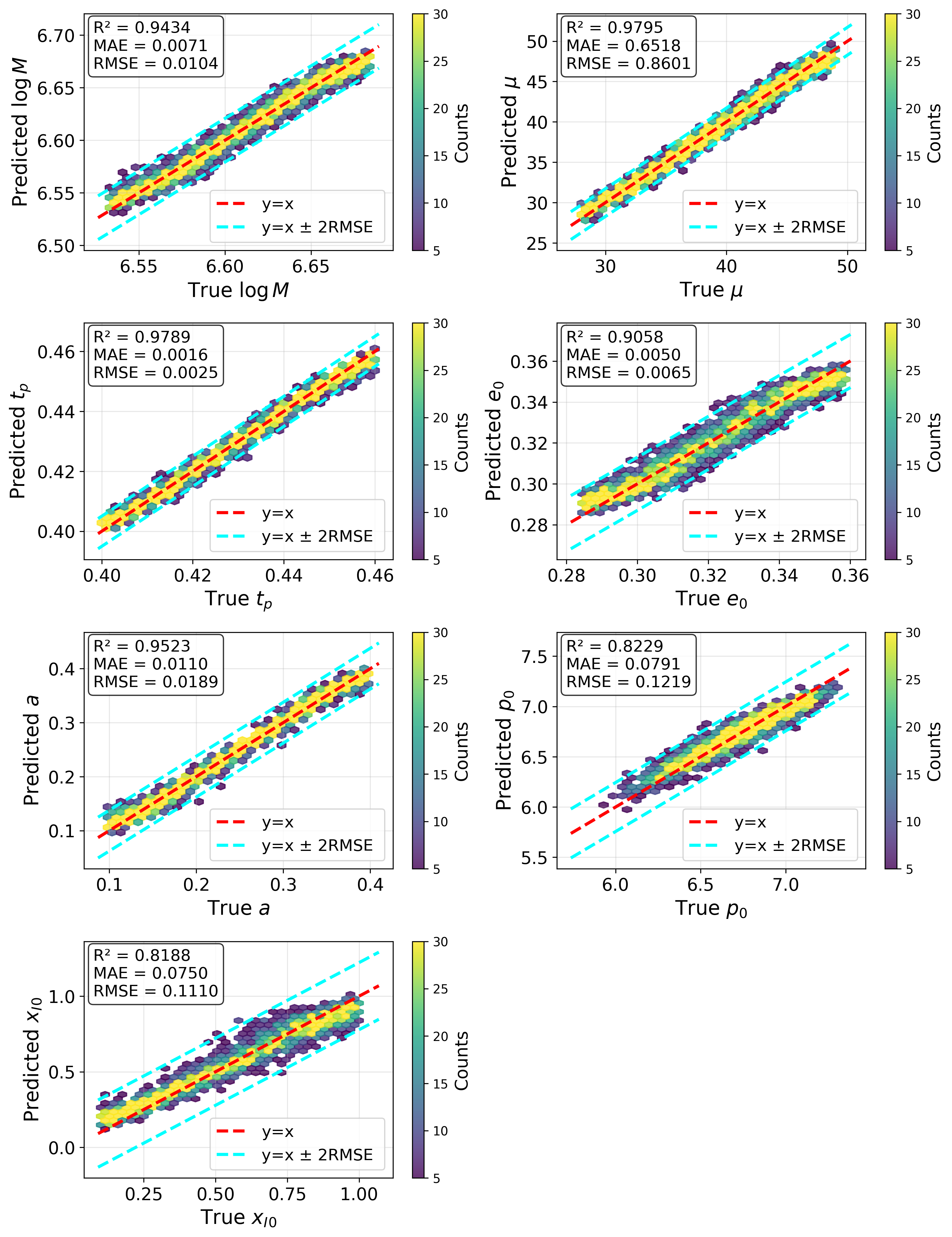}
    \caption{    Recovery performance of our CNN model for the intrinsic parameters $t_p$, $\log M$, $\mu$, $a$, $e_0$, $p_0$, and  $x_{I0}$ (Step 2).
     The predicted values (y-axis) are plotted against the corresponding true values (x-axis). The red dashed line denotes perfect agreement ($y=x$), and the cyan dashed lines indicate the $y=x\  \pm \ 2\ \text{RMSE}$ bounds. Notably, 95.1\% of the data points fall within the $2\,\text{RMSE}$ bounds. The color bar indicates the density of events in each hexagonal bin.
    }
    \label{fig:intri_result_modify}
\end{figure}

Figures~\ref{fig:intri_result} and~\ref{fig:intri_result_modify} illustrate the predicted versus true values for the Step 1 and Step 2 estimations, respectively. A direct comparison highlights a marked improvement in estimation precision following the Step 2 refinement. Specifically, the high-density regions of the hexagonal bins in Step 2 cluster significantly tighter around the line of perfect agreement ($y=x$) compared to Step 1. This observation is quantitatively corroborated by the substantial reduction in both MAE and RMSE across all parameters; for instance, the RMSE for $\mu$ drops from $\sim 5.0$ in Step 1 to $\sim 0.86$ in Step 2. Notably, the parameter $x_{I0}$, which is introduced in Step 2, exhibits a robust alignment along the diagonal with an $R^2$ value of 0.8188. 
These results confirm that narrowing the parameter space and frequency band effectively mitigates parameter degeneracies, thereby yielding highly accurate waveform reconstructions.

 \begin{table}[H]
\footnotesize
\begin{threeparttable}\caption{The injected and recovered parameters of the EMRI signal shown in Fig.~\ref{fig:Unet_extract} across Step 1 and Step 2. The combined SNR of the TDI-A and TDI-E channels is 54.}\label{tab:intrinsic_result_modify}
\doublerulesep 0.1pt \tabcolsep 0.8pt 
\begin{tabular}{ccc|ccc|ccc}
\toprule
&   &  & \multicolumn{3}{c|}{Step 1} & \multicolumn{3}{c}{Step 2} \\
Param. & Prior & $P_{\rm inj}$ & $P_{\rm est}$ & RMSE &  $\frac{4\times \rm{RMSE}}{ \rm{Prior}}$ & $P_{\rm est}$ & RMSE &  $\frac{4\times \rm{RMSE}}{ \rm{Prior}}$ \\  \hline
$\log M$   & [6.0,7.0] & 6.5927 & 6.6084  & 0.0377    & 0.1508 & 6.5792   & 0.0104 & 0.0416 \\
$\mu$      & [20,100]  & 36.1605  & 38.2995  & 5.0599  & 0.2530 & 34.8340  & 0.8601 & 0.0430 \\
$t_{p}$    & [0.4,0.46]& 0.4141  & 0.4145 & 0.0041    & 0.2733  & 0.4157  & 0.0025 &  0.1667\\
$e_0$      & [0.1,0.6] & 0.3288 & 0.3217 & 0.0186     &0.1488   &0.3419 &0.0065 & 0.0520\\
$a$        & [0.1,0.9] & 0.2469  & 0.2481 &  0.0750   & 0.3750 & 0.2252  &0.0189 &0.0945 \\
$p_0$        & [3.8,11.7] & 6.9105  & 6.7359 & 0.4239  &0.2146 & 6.8750  & 0.1219 & 0.0617\\
${x_I}_0$     & [0.1,1.0] &  0.2969   &——     &——      & ——   & 0.2830  &  0.1110 & 0.4933 \\
\bottomrule
\end{tabular}
\end{threeparttable}
\end{table}

\begin{figure}[H] 
\centering
	\includegraphics[width=0.48\textwidth]{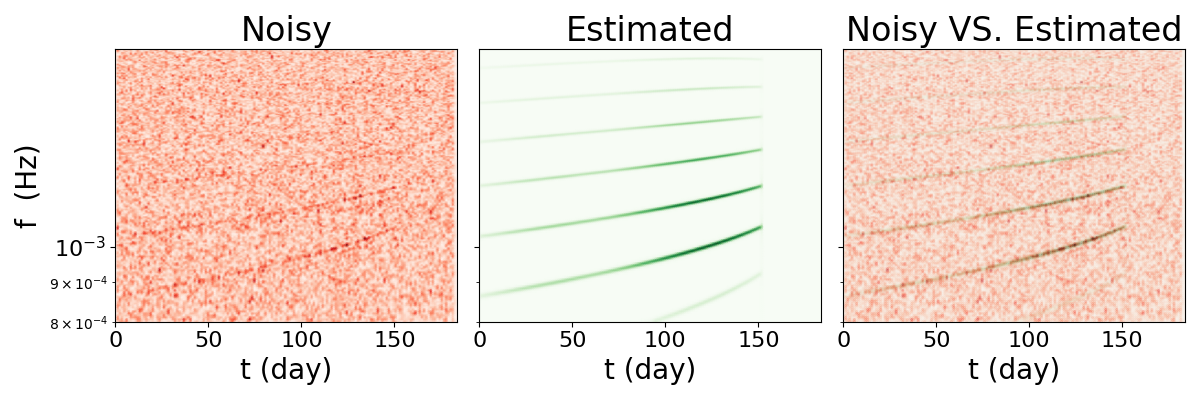}
    \caption{Time-frequency domain comparison between the noisy EMRI signal and the CNN-reconstructed signal (Step 2). 
The intrinsic parameters ($\log M$, $\mu$, $t_p$, $e_0$, $a$, and $x_{I0}$ ) of the recovered waveform are assigned the CNN-predicted values $P_{\rm est}$ from Step 2, with $D_L = 1\,\text{Gpc}$. The remaining parameters are set to the median of their prior distributions.
} 
    \label{fig:our CNN_result_compare_2}
\end{figure}

 Tab.~\ref{tab:intrinsic_result_modify} summarizes the comparison between the injected parameters ($P_{\rm inj}$) and those estimated ($P_{\rm est}$) by our two-step process for the EMRI signal depicted in Fig.~\ref{fig:Unet_extract}. In the table, ``Prior'' indicates the initial search range, while the metric $4\times{\rm RMSE/Prior}$ represents the reduction ratio of the parameter space range, quantifying how many times the search space has been compressed relative to the prior. As evidenced by the results, the RMSEs for all parameters decrease significantly after the Step 2 refinement, leading to much higher reduction ratios, especially for $\mu$. It is particularly noteworthy that while the initial inclination cosine ${x_I}_0$ yields no valid estimate in Step 1 (denoted by ``---''), it is successfully recovered in Step 2, underscoring the critical role of the refinement stage for constraining this specific parameter.

As shown in Fig.~\ref{fig:our CNN_result_compare_2}, the EMRI signal estimated by our CNN in Step 2 exhibits excellent agreement with the noisy signal in the time-frequency domain. The model successfully captures the main frequency evolution track, maintaining high consistency in both the spectral energy distribution and the temporal duration of the signal, which further validates the accuracy of the recovered intrinsic parameters.

Collectively, these results validate the efficacy of our proposed deep-learning-based hierarchical parameter estimation framework. We successfully recover the seven intrinsic parameters of spinning black hole EMRI systems, reducing the search ranges for $\{\log M, a, \mu, e_0, p_0 \}$ and $\{t_p, {x_I}_0 \}$ by factors of $\sim 10^{2}$ and $\sim 10^{1}$, respectively. This corresponds to a total compression of the seven-dimensional parameter space volume by a factor of $\sim 10^{8}$.

\section{\label{sec:conclusion} Conclusions} 

EMRIs are among the primary targets for future space-borne GW detectors. These systems carry rich scientific information with broad applications in astrophysics, cosmology, and fundamental physics. To fully exploit these scientific potentials, precise parameter extraction is essential. However, EMRI parameter estimation remains a highly challenging task due to the complexity of the waveforms and the high dimensionality of the parameter space.
In this work, we develop a deep-learning-based framework to recover the seven intrinsic parameters of EMRI systems involving spinning black holes without relying on detailed priors for extrinsic parameters. Our proposed pipeline consists of three key stages, which effectively address the challenges of high-dimensional parameter estimation.

First, to tackle the issue of heavy detector noise contamination, we employed a U-Net framework to extract signal information from noisy STFT images in the time-frequency domain. By adopting a cosine similarity threshold of 0.8, we achieved a success rate exceeding 80\% for signals with $\rm{SNR} > 40$ and surpassing 95\% for golden EMRI events ($\rm{SNR} > 50$). These results demonstrate that the U-Net provides a robust solution for EMRI signal denoising under moderate-to-high SNR conditions, effectively preserving track continuity and fine-grained features.

Second, leveraging the extracted time-frequency representations, we designed a multi-channel CNN regression framework for coarse intrinsic parameter estimation. A dual attention mechanism (frequency attention and time attention) is further incorporated to refine feature extraction across both time and frequency dimensions. By decoupling these dimensions, the mechanism jointly models long-range temporal evolution and persistent spectral structures, thereby enhancing sensitivity to EMRI time–frequency ridges in noisy data. 
This stage successfully narrows down the search region, yielding a total reduction of the 7-dimensional parameter space by a factor of $\sim 10^{4}$.

Finally, leveraging the coarse estimates from the previous stage, we refine the CNN-based parameter estimation within a narrowed search region. In this refinement stage, the ranges of $\{\log M, a, \mu, t_p, e_0, p_0\}$ and ${x_I}_0$ are further compressed by factors of $\sim 10^{2}$ and $\sim 10^{1}$, respectively. This results in an overall reduction of the 7-dimensional parameter space volume by a factor of $\sim 10^{8}$. Such a dramatic reduction enables rapid pre-screening of the intrinsic parameter space, substantially saving computational cost for subsequent full parameter estimation.

It should be noted that the current work does not account for transient, non-Gaussian noise artifacts or ``glitches''~\cite{2025_EMRI_glitch}, nor does it consider complex astrophysical environmental factors~\cite{yuan2026environmentalimprintsdarkmatter,li2026constraintenvironmentseccentricextrememassratio,dittmann2026extrememassratioinspiralsgaseousdisks}. The noise model employed in this study is relatively ideal, and these limitations will be addressed in future improvements. Furthermore, our U-Net-based signal extraction performs well for signals with SNR greater than 50, but its effectiveness for low-SNR scenarios remains to be improved. Future work will focus on enhancing the robustness of the model under more realistic noise conditions and extending its applicability to lower SNR regimes.

 The remaining extrinsic parameters mainly affect the amplitudes of the STFT maps, which will be considered in future work. We plan to develop a hybrid framework that combines traditional matched filtering techniques~\cite{Dhurkunde_2022_matched_filtering,guo2026longalgorithmbasedquantumsearch,Usman_2016_pycbc,Allen_2012} with deep learning methods~\cite{lipman2023flowmatchinggenerativemodeling,vu2026selfforcecalculationsnumericalrelativity,2024_LB_EMRI_PE} to achieve full parameter estimation. Specifically, we will integrate optimization and sampling algorithms such as Particle Swarm Optimization (PSO)~\cite{PSO_1995,pyswarmsJOSS2018,innocente2021_PSO,2025_ZXB_EMRI_bumpy,zou2026PSO}, Nested Sampling~\cite{2004Nested-Sampling,2006Nested-Sampling,2019Dynamic-nested-sampling,2020DYNESTY,2025Dynesty}, Markov Chain Monte Carlo (MCMC)~\cite{Robert_2011_MCMC, Foreman_Mackey_2013_MCMC,hill2026exactglobalmcmcdenoising}, and Flow Matching ~\cite{lipman2023flowmatchinggenerativemodeling,vu2026selfforcecalculationsnumericalrelativity,2024_LB_EMRI_PE} to efficiently explore the complete parameter space and improve the overall estimation accuracy.

\Acknowledgements{This work was supported by the National Science and Technology Major Project (No. 2024ZD1100601), the National Key R$\&$D Program of China (Grant No. 2021YFC2203002), the National Natural Science Foundation of China (Grant No. 12605093, 12473075, 12173071). This work made use of the High Performance Computing Resource in the Core Facility for Advanced Research Computing at Shanghai Astronomical Observatory.}


\InterestConflict{The authors declare that they have no conflict of interest.}

\bibliography{ref}
\bibliographystyle{scpma}

\end{multicols}
\end{document}